\documentclass[amsmath,amssymb,aps,prd,11pt,tightenlines,superscriptaddress, 
nofootinbib,preprintnumbers,notitlepage]{revtex4-1}
\usepackage{graphicx}
\usepackage{dcolumn}
\usepackage{bm}
\usepackage{amssymb}
\usepackage{amsmath}
\usepackage{mathrsfs}
\usepackage{graphicx,color} 
\usepackage[english]{babel}     
\usepackage{graphicx,wrapfig,lipsum}      
\usepackage{multimedia}    
\usepackage[mathscr]{eucal}  
\usepackage{bm}              
\usepackage{amssymb}            
\usepackage{amsmath}     
\usepackage{mathrsfs}  
\usepackage[mathscr]{eucal}      
\usepackage[normalem]{ulem}
  
\begin{document}     
    
\title{BPS strings and the stability of the asymptotic Casimir law in adjoint flavor-symmetric YMH models}  
\author{David R. Junior}
\author{Luis E. Oxman}
\author{Gustavo M. Sim\~oes}  
  
\affiliation{
Instituto de F\'\i sica, Universidade Federal Fluminense, 24210-346 Niter\'oi - RJ, Brasil.} 
\date{\today} 
 
\begin{abstract}     
We investigate an effective flavor-symmetric Yang-Mills-Higgs model with $N^2-1$ adjoint scalar fields. We find a set of BPS equations 
that provide vortex solutions and calculate their energies for arbitrary representations. 
We show that, for a given N-ality $k$, the energy of the corresponding antisymmetric representation is the lowest. 
This completes the proof that this model is able to reproduce a Casimir law for the string tension at asymptotic distances. 
\end{abstract}   
 
\maketitle
\section{Introduction}
The chromoelectric flux tube between external quarks in $SU(N)$ Yang-Mills (YM) theory \cite{Bali:2000gf,Bali2000CS,Luscher:2002qv,Teper2004,Cosmai2017,Yanagihara2019210}  has many interesting properties. At intermediate distances,  the lattice string tension $\sigma_{\rm I}({\rm D})$, derived from the Wilson loop average 
$\langle W_{\mathcal C} \rangle$\footnote{ $\mathcal{C}$ is the closed worldline associated with the external quark/antiquark pair.}, scales with the quadratic Casimir $C_2({\rm D})$ of the $SU(N)$ quark representation $D(\cdot)$, see Ref. \cite{Bali2000CS}. That is, 
\begin{equation}
   \frac{\sigma_{\rm I}({\rm D})}{\sigma_{\rm I}({\rm F})} = \frac{ C_2({\rm D})}{C_2({\rm F})}\;,
   \label{inter}
\end{equation}  
where F stands for the fundamental representation. 
In this work, we will be mainly interested in the behavior at asymptotic distances, where the string tension is known to depend only on the $N$-ality of  $D(\cdot)$  \cite{KRATOCHVILA2003103}. The latter is given by an integer $k$ (modulo $N$) that dictates how the center of $SU(N)$,
\begin{equation}
    Z(N) = \left\{z\, \mathbb{I}_N\vert \, z\in \mathbb{C},\,z^N=1\right\}\;,
\end{equation}
 is realized. Namely 
\begin{equation}
    {\rm D}(z\, \mathbb{I}_N) = z^k \, \mathbb{I}_{\mathcal{D}} \;,
\end{equation}
where $\mathbb{I}_{\mathcal{D}}$ is a $\mathcal{D} \times \mathcal{D}$ identity matrix and $\mathcal{D}$ is the dimension of ${\rm D}$. 

A possible approach to capture the physics at asymptotic distances is to look for effective Yang-Mills-Higgs (YMH) models that accomodate $N$-ality as due to an $SU(N) \to Z(N)$ spontaneous symmetry breaking (SSB) pattern, which leads to the formation of $Z(N)$ strings \cite{Oxman:2012ej,oxmangustavo}.  In this scenario, the quarks are represented by external monopole/antimonopole pairs carrying the charges (weights) that characterize ${\rm D}(\cdot)$ \cite{article}.  
In Ref. \cite{oxman4d} (see also \cite{Oxman:2019c+}), the pure Yang-Mills sector of these models was associated to  the continuum limit of an effective Wilson action with frustration. The latter generates an average of center elements which depends on the linking number between the external quark worldline $\mathcal{C}$ and plaquette configurations distributed on closed surfaces. These configurations were thus identified with an ensemble of center-vortex worldsurfaces, which are  quantum variables extensively explored in the lattice as a source for confinement \cite{DelDebbio:1996lih,Langfeld:1997jx,DelDebbio:1998luz,Faber:1997rp,deForcrand:1999our,Ambjorn:1999ym,Engelhardt:1999fd,Engelhardt:1999xw,Bertle:2001xd,Reinhardt:2001kf,Gattnar:2004gx}. The possibility of nonoriented center vortices, where the $\mathfrak{su}(N)$ Lie algebra orientation changes at some worldlines on the worldsurfaces, was also observed in lattice simulations, and is believed to play a relevant role for confinement \cite{KRONFELD1987516,chernodub,PhysRevD.61.034503}. In Ref. \cite{oxman4d}, this type of nonoriented object was introduced by means of an ensemble of adjoint dual Wilson loops.  Moreover, in the continuum, this sector induces a set of effective  adjoint Higgs fields, while possible correlations between adjoint dual Wilson lines were related to effective Higgs interactions. Then, among the 
$SU(N) \to Z(N)$ models, an interesting possibility is the one introduced in Ref. \cite{Oxman:2012ej}, as it contains $N^2-1$ adjoint Higgs fields, and displays flavor symmetry. 

Besides $N$-ality, an effective description should also explain the particular scaling of the asymptotic string tension $\sigma({\rm D})$. There are two main possibilities consistent with the  lattice data \cite{Lucini_2004}. One of them is the sine-law
\begin{equation}
   \frac{\sigma({\rm D})}{\sigma({\rm F})} = \frac{ \sin{(k\pi/N)}}{\sin{(\pi/N)}}\;.
\end{equation} 
The other one is an extension to the asymptotic region of the behavior in Eq. \eqref{inter},  but replacing $C_2({\rm D})$ by the lowest quadratic Casimir among representations with the same 
$N$-ality than ${\rm D(\cdot)}$. The latter is given by the $k$-Antisymmetric $(k\text{-}{\rm A})$ irrep.  
Interestingly, in the adjoint flavor-symmetric model,  the tension of the infinite $k\text{-}{\rm A}$ string  scales with the quadratic Casimir \cite{oxmangustavo}, which is compatible with the second possibility. In this context, in order to estalish the asymptotic Casimir scaling law one must also show that this is the lowest tension among the irreps with $N$-ality $k$. In that case,  $k\text{-}{\rm A}$ strings would be settled as the stable  confining states. This is one of the properties we will be able to address exactly in this work. For this aim, we need an analysis of the field equations for any representation ${\rm D}(\cdot)$ of $SU(N)$, which was still lacking.  
Here,  we will show that there is a point in parameter space where the
complicated set of second order equations can be reduced to a set of first order BPS equations. 
In flavor-symmetric models, this reduction was shown in Refs. \cite{ETO200898, PhysRevD.69.065003,PhysRevD.70.025013} when the Higgs fields are in the fundamental representation,  and in Ref. \cite{PhysRevD.95.025001}, only for $SU(2)$, when the Higgs fields are in the adjoint. 
At this point, we will  close an ansatz for a string carrying any weight of $SU(N)$ showing that, for a given $N$-ality $k$, the tension corresponding to the k-antisymmetric representation is indeed the lowest.

\section{Previous results}

The flavor-symmetric effective model with adjoint Higgs fields $\psi_I$, which take values in the $\mathfrak{su}(N)$ Lie algebra, is given by\footnote{Throughout this work, we use Euclidean metric.}   \cite{Oxman:2012ej} 
\begin{subequations}
\label{action}
\begin{align}
 S=\int d^4x \left( -\frac{1}{4} \langle F_{\mu\nu},F_{\mu\nu}\rangle + \frac{1}{2}\langle D_\mu \psi_I, D_\mu \psi_I \rangle - V_{\rm H}(\psi) \right)  \;,\\    
F_{\mu\nu} =\frac{i}{g} \left[D_\mu,D_\nu\right]\makebox[.5in]{,}  D_\mu = \partial_\mu - ig[A_\mu,\;]=\partial_\mu +g A_\mu\wedge \;.   
\end{align}
\end{subequations}
The number of flavors $I=1,\dots,N^2-1$ equals the dimension of $\mathfrak{su}(N)$.    Under a gauge transformation $U\in SU(N)$, we have 
\begin{subequations}
\label{gaugetransformation}
\begin{align}
       A_\mu\rightarrow UA_\mu U^{-1} + \frac{i}{g} U \partial_\mu U^{-1}\label{gaugetransformation1}
       \makebox[.5in]{,}   \psi_I\rightarrow U\psi_I U^{-1}\;.
\end{align}
\end{subequations}
The potential was set as
\begin{equation}
\label{potentialhiggs}
V_{\rm H}(\psi) = c+\frac{\mu^2}{2}\langle \psi_A,\psi_A\rangle + \frac{\kappa}{3}f_{ABC}\langle \psi_A \wedge \psi_B, \psi_C\rangle + \frac{\lambda}{4}\langle \psi_A\wedge \psi_B\rangle^2\;,
\end{equation}
which leads to the classical vacua ($S \in SU(N)$)
\begin{subequations}
\begin{align}
 A_\mu = \frac{i}{g}S\partial_\mu S^{-1}    \makebox[.5in]{,}   \psi_A = v S T_A S^{-1}   \;.
\end{align}
\end{subequations}
 Here, $T_A$ and $f_{ABC}$  are the $\mathfrak{su}(N)$ Lie basis and structure constants, respectively.  Throughout this work, we shall also separate the color and flavor indices into Cartan $q= 1, \dots, N-1$  and off-diagonal $\alpha$, $\bar{\alpha}$ labels. The elements $T_q$ form a maximal commuting set, while the remaining elements are defined in terms of root vectors $E_{\pm \alpha}$ 
 \begin{equation}
    T_\alpha = \frac{E_\alpha + E_{-\alpha}}{\sqrt{2}} \makebox[.5in]{,}  T_{\bar{\alpha}}=\frac{E_\alpha - E_{-\alpha}}{\sqrt{2}i}\;,
    \label{hermit}
\end{equation} 
where $\alpha$ is a positive root of $\mathfrak{su}(N)$. For the notation and conventions, see Appendix \ref{appA}.
 
As the only transformation that leaves a Higgs field vacuum configuration invariant  is $U = z \, \mathbb{I}_N$,
the system undergoes $SU(N)\rightarrow Z(N)$  SSB. Consequently, the vortex solutions to the static field equations 
\begin{subequations}    
\label{Modeleq}
    \begin{align}
        D_j F_{ij} &=gD_ i\psi_A\wedge\psi_A\;,\label{gaugeeq}\\
        D_iD_i\psi_A&= \frac{\delta V_H}{\delta \psi_A}\label{higgseq}\;,
    \end{align}
\end{subequations}
are topologically stable due to the nontrivial first homotopy group of the associated vacua manifold  $\mathcal{M} = \frac{SU(N)}{Z(N)}$, $\Pi_1\left(\mathcal{M}\right) = Z(N)$.
To find these solutions, the ansatz
\begin{gather}
\label{ansatz}
    A_0=0  \makebox[.5in]{,}  A_i = S\mathcal{A}_iS^{-1} + \frac{i}{g}S\partial_i S^{-1}    \makebox[.5in]{,} \psi_A = h_{AB} S T_A S^{-1} \makebox[.5in]{,} S = e^{i\varphi\beta\cdot T}\;
\end{gather} 
was used. For  infinitely long strings with cylindrical symmetry, the profiles $a$ and $h_{AB}$ can be taken as functions of $\rho$ alone, with $(\rho,\,\varphi,\,z)$ being cylindrical coordinates. The vortex charge is defined by the magnetic weight $\beta =2N \omega$, with $\omega$ being the highest weight of the representation ${\rm D}(\cdot)$. Here, we used the notation $\beta\cdot T = \beta|_q T_q$, where $\beta|_q $ is the $q$-th component of the $(N-1)$-tuple $\beta$.   For the various definitions and properties, see Appendix \ref{wei}. In Ref. \cite{oxmangustavo}, considerig $
 \mathcal{A}_i = (a/g) \partial_i\varphi\beta\cdot T $, we obtained vortex solutions for  the $k$-A and $k$-Symmetric  ($k$-S) representations. In this work, using $\mathcal{A}_i$ along a general Cartan direction (cf. Sec. \ref{ans-g}), we shall be able to accomodate a vortex for a general ${\rm D}(\cdot)$.  
In terms of the Cartan-Weyl sectors, the anstaz has the simpler structure:
\begin{equation}
\label{ansatz2}
    \psi_\alpha = h_\alpha S T_\alpha S^{-1} \makebox[.3in]{,} \psi_{\bar{\alpha}} = h_\alpha S T_{\bar{\alpha}} S^{-1} \makebox[.3in]{,} \psi_q = h_{qp}S T_p S^{-1}\;.
\end{equation}
In order for the gauge and $\psi_\alpha,\psi_{\bar{\alpha}}$ fields, with $\alpha\cdot\beta\ne 0$, to be
 well-defined along the $z$ axis, we imposed the regularity conditions  
\begin{subequations}
\begin{align}
    a(0) &= 0\;,\\ 
    h_{\alpha}(0)&=0 \makebox[.8in]{\rm when}  \alpha\cdot \beta \ne 0\;.
\end{align}
\end{subequations}
 In this regard, note that
\begin{subequations}
\begin{align}
    S T_\alpha S^{-1} &= \cos \left(\varphi \beta\cdot \alpha \right) T_\alpha + \sin\left(\varphi \beta\cdot \alpha \right) T_{\bar{\alpha}}\;,\\
    S T_{\bar{\alpha}} S^{-1} &= \cos \left(\varphi \beta\cdot \alpha \right) T_{\bar{\alpha}} - \sin\left(\varphi \beta\cdot \alpha \right) T_\alpha\;.
\end{align}
\label{rota}
\end{subequations}    
When $\mu^2=0$, the solution for the fields with no regularity conditions at $\rho=0$ is frozen everywhere at the vacuum value:
\begin{equation}
    \psi_q = v T_q \makebox[.3in]{,} \psi_\alpha= v T_\alpha \makebox[.8in]{\rm when} \alpha\cdot \beta=0\;.
\end{equation}
This led to the following asymptotic exact behavior of the string tension for the $k$-A representation  
\begin{equation}
    \frac{\sigma(k\text{-}{\rm A})}{\sigma({\rm F})} = \frac{k(N-k)}{N-1} = \frac{C_2(k\text{-}{\rm A} )}{C_2({\rm F})}\;,
\end{equation}
 This agrees with the large distance behavior of the Wilson loop \cite{Teper2004}.   It is trivial to extend the discussion of Ref. \cite{oxmangustavo} to the $k$-S irrep. In this case, the model is equivalent to a Ginzburg-Landau theory with winding number $k$. Then, at the BPS point $\lambda=g^2$ of the Abelianized $\mu^2 = 0$ model, we have
\begin{equation}
    \frac{\sigma(k\text{-}\rm{S})}{\sigma(\rm{F})}=k > \frac{k(N-k)}{N-1}=\frac{\sigma(k\text{-}\rm{A})}{\sigma(\rm{F})}\;,
\end{equation}
for $k>1$. Then, when a $k$-S string is long enough, it is energetically favorable to create valence gluon excitations around the quark sources to produce a $k$-A string.

\section{BPS equations} 

In the Nielsen-Olesen model governed by the action ($D_\mu = \partial_\mu -ig A_\mu$, $\phi \in \mathbb{C}$) 
\begin{equation}
    S_{\rm Abe} = \int d^4x \Big(  -\frac{1}{4}F_{\mu\nu} F_{\mu\nu} + \frac{1}{2}D_\mu \phi D_\mu\phi -\frac{\lambda}{8}(\phi\phi^*-v^2)^2 \Big)  \;,
\end{equation}
 when $\lambda< g^2$, a single vortex with higher winding number $n$ is energetically more favorable than $n$ separated vortices with winding number $1$. When $\lambda>g^2$, the situation is reversed. At $\lambda=g^2$, also known as the BPS point, the vortices do not interact, as the energy of any configuration with winding number $n$ is given by
\begin{equation}
    E = g v^2\int d^3x\, B_3 = 2\pi v^2 n\;.
\end{equation}
In this Abelian setting,  the equations of motion at the BPS point can be reduced to be first order
\begin{equation}
\label{BPSAbelian}
    D_+ \phi = 0\;,\; B_3 = \frac{g}{2} (v^2-\phi\phi^*)\;,\; B_1=B_2=0\;,
\end{equation} 
where $D_\pm = D_1\pm i D_2$.  
For a detailed discussion on this topic, see Ref. \cite{manton2004topological}. In the non Abelian context, this type of BPS point is known to occur in flavor-symmetric $SU(N) \to Z(N)$ models constructed in terms of $N$ Higgs fields in the fundamental representation \cite{ETO200898, PhysRevD.69.065003, Hanany_2004}. In this section, we will show that there is a set of BPS equations that provide solutions to the flavor-symmetric  $SU(N) \to Z(N)$ model formed by $N^2-1$ adjoint Higgs fields, at 
 $\mu^2=0$ and $\lambda=g^2$ (cf. Eqs. \eqref{action}, \eqref{potentialhiggs}, \eqref{Modeleq}).  Moreover, we will show that these equations can be closed with an ansatz that accommodates center vortices carrying the weights of any $SU(N)$ group representation.  

Initially, for every pair $\psi_\alpha,\psi_{\bar{\alpha}}$, with $\alpha>0$, we define 
\begin{equation}
\label{Zetadefinition}
    \zeta_{\alpha} = \frac{\psi_\alpha + i\psi_{\bar{\alpha}}}{\sqrt{2}}\;,
\end{equation}
which is in  the complexified $\mathfrak{su}(N)$ Lie algebra ($\alpha$ is a positive root). We shall consider configurations for an infinite static vortex. Because of translation symmetry along the $x^3$-direction, we require  
\begin{align}  
 B_1=B_2=0 \makebox[.5in]{,}     D_3\psi_A&=0 \;. \label{BPStrivial}
\end{align} 
Next, motivated by the BPS equations in Refs. \cite{ETO200898,Kneipp200469,PhysRevD.69.065003} involving Higgs fields transforming in the fundamental and adjoint representations, for the field-dependence transverse to the string we propose the first-order equations   
\begin{subequations}
\label{BPSeq}
\begin{gather}
     D_+\zeta_\alpha = 0 \makebox[.15in]{}\Leftrightarrow\makebox[.15in]{} D_-\zeta^\dagger_{\alpha}=0 \makebox[.3in],{}  D_1\psi_q = D_2 \psi_q =0 \;, \label{BPSZeta+}\\
    B_3 =  g \sum\limits_{\alpha>0}\left(v \alpha|_q \psi_q  -[ \zeta_\alpha,\zeta_\alpha^\dagger]
   \right)\;. \label{BPSMagnetic} 
    \end{gather} 
\end{subequations}
In terms of the original fields, we can also write
\begin{gather} 
   D_{\pm}\psi_{\alpha} =\mp i D_{\pm}\psi_{\bar{\alpha}}\;,\label{BPSpsialpha}\\
    B_3 = g \sum\limits_{\alpha>0}\left(v \alpha|_q \psi_q - \psi_\alpha\wedge\psi_{\bar{\alpha}}\right)\label{BPSMagnetic2}\;. 
\end{gather}  

\subsection{The ansatz}   
\label{ans-g}

Regarding the ansatz, we shall use  Eqs. \eqref{ansatz} and \eqref{ansatz2}, with  $\mathcal{A}_i $ being a general field in the Cartan subalgebra 
$\mathfrak{C}$, not necessarily proportional to $\beta\cdot T$,
\begin{equation}
    \mathcal{A}_i = \sum\limits_{l=1}^{N-1}\frac{a_l -d_l}{g}\partial_i\varphi\beta^{l-\rm{A}}\cdot T\;,
    \label{Car} 
\end{equation} 
where $\beta^{(l)} = 2N \Lambda^{l-\text{A}}$ and $\Lambda^{l-\text{A}}$, $l=1, \dots, N-1$ are the antisymmetric (fundamental) weights, which provide a basis $\beta^{(l)}\cdot T   $ for $\mathfrak{C}$. The Dynkin numbers $d_l$ are the positive integer coefficients obtained when expressing $\beta$ as a linear combination of $\beta^{l-\rm{A}}$. The profiles $a_l$ must obey the boundary conditions
\begin{equation}
    a_l(0)=0\makebox[.5in]{,}a_l(\infty)=d_l\;.
\end{equation}
The first guarantees the a finite action density and a well-defined strength field along the vortex core while the second ensures that the gauge field is a pure gauge, cf. \eqref{ansatz},
\begin{equation}
\label{AsymptoticGauge}
    A_i\rightarrow \frac{\partial_i\varphi}{g}\beta\cdot T\makebox[.5in]{, when } \rho\rightarrow\infty.
\end{equation}

From this ansatz, it also follows that   $D_i \psi_q = \partial_i \psi_q$ and, from Eqs. \eqref{BPStrivial}, \eqref{BPSZeta+}, that the fields $\psi_q$ must be homogeneous. We shall take  $\psi_q \equiv  v T_q$

 Notice that Eq. \eqref{BPSZeta+} leads to
\begin{equation}
    D_{+}\left[\zeta_\alpha,\zeta_{\alpha'}\right] = \left[D_+\zeta_\alpha,\zeta_{\alpha'}\right] + \left[\zeta_\alpha,D_+\zeta_{\alpha'}\right]=0\;,
\end{equation}
if both $\alpha$ and $\alpha'$ are positive roots. This suggests that $\left[\zeta_\alpha,\zeta_{\alpha'}\right]$ is proportional to another $\zeta_{\alpha''}$. In addition,  the boundary conditions imply
\begin{equation}
    \left[\zeta_\alpha,\zeta_{\alpha'}\right]\rightarrow v^2 \mathcal{N}_{\alpha,\alpha'} \left[E_{\alpha},E_{\alpha'}\right] = v^2 \mathcal{N}_{\alpha,\alpha'} E_{\alpha+\alpha'}\text{ when } \rho\rightarrow \infty\;.
\end{equation}
Then, it is natural to assume
\begin{equation}
\label{CommutatorZetas}
    \left[\zeta_{\alpha},\zeta_{\alpha'}\right] = v \mathcal{N}_{\alpha,\alpha'} \zeta_{\alpha+\alpha'}
    \;.
\end{equation}    
Regarding this proposal, it is important to check if it is consistent with the regularity conditions
at $\rho=0$. Fortunately, when both $\alpha,\alpha'$ are positive roots, these equations are always  consistent.

If  $\alpha\cdot\beta\ne0$,  because of the ansatz \eqref{ansatz2} and Eq. \eqref{rota}, we must impose $\zeta_\alpha(\rho\rightarrow0)=0$.  These conditions are compatible as the highest weight is always a positive integer linear combination of fundamental weights (see App. 
 \ref{wei}). In addition,  the inner product between a  fundamental weight and a positive root is positive. Therefore,  if  $\beta \cdot \alpha \neq 0$ or $\beta \cdot \alpha' \neq 0 $, then $\beta \cdot (\alpha+\alpha') \neq 0$. In this case,  
 to avoid the defect in Eq. \eqref{rota}, $\zeta_{\alpha+\alpha'}$  will be zero at $\rho=0$, in accordance with the regularity condition on at least one of the factors in the left-hand side of Eq. \eqref{CommutatorZetas}. 
On the other hand, when both $\beta \cdot \alpha = 0$ and $\beta \cdot \alpha' =0$, the associated basis elements do not rotate so $\psi_\alpha$, $\psi_{\bar{\alpha}}$, $\psi_{\alpha'}$, $\psi_{\bar{\alpha}'}$ are not fixed at the origin. In this case, just like $\psi_q$, it holds that $D_i\psi_\alpha=\partial_i\psi_\alpha$. For this reason, when $\beta\cdot\alpha=0$ we will assume  $\psi_{\alpha} = vT_{\alpha}$, $\psi_{\bar{\alpha}}= v T_{\bar{\alpha}}$. Consequently, Eq. \eqref{CommutatorZetas} also holds in this case, as it simply follows from  the commutation relations between $E_\alpha$ and $E_{\alpha'}$. Moreover, it is not difficult to check that this solves the equations for $\psi_\alpha$ when $T_\alpha$ and $T_{\bar{\alpha}}$ do not rotate.  
 
 \subsection{Reduced scalar BPS equations}

 Notice that
\begin{eqnarray}  
    D_+(A) \zeta_{\alpha} &=& S D_+(\mathcal{A}) (h_\alpha E_\alpha) S^{-1} = \Big(\partial_+h_{\alpha}-i \partial_+\varphi h_{\alpha} \sum\limits_{l=1}^{N-1}(a_l-d_l) \alpha\cdot \beta^{l-\rm{A}}\Big) SE_{\alpha}S^{-1}\;,\\
    B^3 &=& \sum\limits_{l=1}^{N-1}\frac{1}{g\rho}\frac{\partial a_l}{\partial\rho}\beta^{l-\rm{A}}\cdot T = g \sum\limits_{\alpha>0}v^2 \alpha\cdot T - \psi_\alpha\wedge\psi_{\bar{\alpha}} = g \sum\limits_{\alpha>0}(v^2- h_\alpha^2) S \alpha\cdot T S^{-1}\;.
\end{eqnarray}
These two relations imply the BPS equations for the the gauge and Higgs profiles 
\begin{subequations}
\label{BPSProfileEq}
    \begin{align}
        \partial_+ \ln h_\alpha &= i \partial_+\varphi  \sum\limits_{l=1}^{N-1}(a_l-d_l) \alpha\cdot \beta^{l-\rm{A}}\;,\label{BPSProfileEqHiggs}\\
        \frac{1}{\rho} \frac{\partial a_l}{\partial \rho} &= g^2 \sum\limits_{\alpha>0}(v^2-h_\alpha^2) \alpha\cdot \alpha^{(l)}\;.\label{BPSProfileEqGauge}
    \end{align}
\end{subequations} 
Here, we used the well-known property involving the fundamental weights and the simple roots $\alpha^{(p)} = \omega_p-\omega_{p+1}$:
\begin{equation}
    \alpha^{(p)}\cdot\beta^{l-\rm{A}}=\delta^{pq}\;.
\end{equation} 
 We have already discussed the property $\zeta_{\alpha}\wedge\zeta_{\alpha'} = v \zeta_{\alpha+\alpha'}$. Naturally, this leads to $h_{\alpha}h_{\alpha'} = v h_{\alpha+\alpha'}$, which is consistent with Eq. \eqref{BPSProfileEqHiggs}. Furthermore, as a general root can be written as a positive sum of simple roots with unit coefficients, the profiles $h_{\alpha^{(p)}}$  associated with simple roots, which satisfy 
    \begin{align}
        \partial_+ \ln h_{\alpha^{(p)}} &= i \partial_+\varphi  (a_p-d_p) \;,
    \end{align}
can be used to generate all the others.

\section{Making contact with the $SU(N) \to Z(N)$ model} 

\subsection{The gauge-field equations} 

From Eqs. \eqref{BPStrivial}, \eqref{BPSMagnetic}, recalling that
\begin{equation}
    B_i = \frac{1}{2} \varepsilon_{ijk}F_{jk} \makebox[.5in]{,}   F_{ij} =\varepsilon_{ijk} B_k\;,
\end{equation}
we can imply 
\begin{eqnarray}
     D_j F_{ij} &=& \varepsilon_{ijk} D_j B_k =-g \varepsilon_{ij3} D_j (\psi_\alpha\wedge\psi_{\bar{\alpha}})\;. 
\end{eqnarray}
If we take $i=1$ and use the BPS equation  for $\psi_\alpha$, $\psi_{\bar{\alpha}}$, we  get 
\begin{eqnarray}
    D_jF_{1j} &=& -g D_2(\psi_\alpha\wedge\psi_{\bar{\alpha}})=-gD_2\psi_\alpha\wedge\psi_{\bar{\alpha}}-g\psi_\alpha\wedge D_2\psi_{\bar{\alpha}}\notag\\
    &=&\frac{ig}{2}\left(D_+\psi_\alpha\wedge\psi_{\bar{\alpha}}-D_-\psi_\alpha\wedge\psi_{\bar{\alpha}}+\psi_\alpha\wedge D_+\psi_{\bar{\alpha}}-\psi_\alpha\wedge D_-\psi_{\bar{\alpha}}\right)\notag\\
    &=&\frac{ig}{2}\left(-iD_+\psi_{\bar{\alpha}}\wedge\psi_{\bar{\alpha}}-iD_-\psi_{\bar{\alpha}}\wedge\psi_{\bar{\alpha}}+i\psi_\alpha\wedge D_+\psi_\alpha+i\psi_\alpha\wedge D_-\psi_\alpha\right)\notag\\
    &=&-g\left(\psi_\alpha\wedge\frac{D_++D_-}{2}\psi_\alpha+\psi_{\bar{\alpha}}\wedge\frac{D_++D_-}{2}\psi_{\bar{\alpha}}\right) =g D_1\psi_A\wedge\psi_A\;.
\end{eqnarray}
This is nothing but the component $i=1$ of Eq. (\ref{gaugeeq}). A similar calculation can be done for $i=2$, while $i=3$ is trivially satisfied.

\subsection{The Higgs-field equations}

\subsubsection{Cartan sector} 
Now,  to  make  contact with the solutions to the Higgs-field equations \eqref{higgseq}, we have to look for a Higgs potential $V_{\rm H}$ that is compatible with the BPS equations. In particular, Eqs. \eqref{BPStrivial}, \eqref{BPSZeta+}  imply $D_i D^i \psi_q =0$, so that $V_{\rm H}$ must imply  
\begin{equation}
    \frac{\delta V_H}{\delta \psi_q} = 0 
\end{equation}
on the ansatz   given
 in Eqs. \eqref{ansatz}, \eqref{ansatz2} and \eqref{Car}, which closes the BPS equations.  In what follows, we will see that this happens when it is given by Eq. \eqref{potentialhiggs} with $\mu^2=0$ and $\lambda=g^2$. 
In this case, 
\begin{equation}
\label{Force} 
    \frac{\delta V_H}{\delta \psi_A} = \lambda\psi_B\wedge(\psi_A\wedge\psi_B-vf_{ABC}\psi_C)\;, 
\end{equation}
where $v=-\frac{\kappa}{\lambda}$. Indeed, applying the same ansatz, we get
\begin{eqnarray}
 \frac{\delta V_H}{\delta \psi_q} &=&  \lambda\sum\limits_{\alpha>0}\psi_\alpha\wedge(\psi_q\wedge\psi_\alpha-vf_{q\alpha \bar{\alpha}}\psi_{\bar{\alpha}})+\psi_{\bar{\alpha}}\wedge(\psi_q\wedge\psi_{\bar{\alpha}}-vf_{q\bar{\alpha} \alpha}\psi_\alpha)\notag\\
    &=&\lambda v\sum\limits_{\alpha>0} \left(h_\alpha S T_\alpha S^{-1}\right)\wedge \left(\alpha\vert_q h_\alpha ST_{\bar{\alpha}}S^{-1}-\alpha\vert_q h_{\alpha} S T_{\bar{\alpha}}S^{-1}\right) = 0\;. 
\end{eqnarray}

\subsubsection{Off-diagonal sector}

Let us now analyze the equations for fields labelled by roots. The BPS equations lead to
\begin{eqnarray}
\label{BPSpsialphapredic}
    D^2 \zeta_\alpha &=& D_-D_+ \zeta_\alpha - g [B_3,\zeta_\alpha]=g^2\sum\limits_{\alpha'>0}\left[ [\zeta_{\alpha'},\zeta_{\alpha'}^\dagger]-v^2\alpha'\cdot T ,\zeta_\alpha\right]   \;.
\end{eqnarray}
The sum over $\alpha'$ involves all positive roots, including $\alpha$. On the other hand, according to the equations of the model, we have
\begin{align}
    D^2\zeta_\alpha =F_\alpha \makebox[.5in]{,}F_\alpha = \frac{1}{\sqrt{2}} \left(\frac{\delta V}{\delta \psi_\alpha}+i\frac{\delta V}{\delta \psi_{\bar{\alpha}}}\right)  \;.
\end{align}
In view of Eq. \eqref{Force}, $F_\alpha$ receives contributions from the index types $B=q,\alpha, \bar{\alpha},\gamma, \bar{\gamma}$ where $\gamma >0$ is a root different from $\alpha$.  The partial contribution originated from the Cartan labels $B=q$ is given by
\begin{equation} 
    F_\alpha^{(B=q)} = \frac{\lambda}{\sqrt{2}}\psi_q\wedge\left(\psi_\alpha\wedge\psi_q-vf_{\alpha q \bar{\alpha}}\psi_{\bar{\alpha}}+ i\psi_{\bar{\alpha}}\wedge\psi_q-ivf_{\bar{\alpha} q \alpha}\psi_\alpha \right) \;.
\end{equation} 
Using the ansatz equations \eqref{ansatz}, \eqref{ansatz2}, and also $\psi_q = v T_q$, we have
\begin{subequations}
\begin{align}
    \psi_\alpha \wedge \psi_q =v f_{\alpha q \bar{\alpha}}\psi_{\bar{\alpha}}\;,\\
    \psi_{\bar{\alpha}} \wedge \psi_q =v f_{\bar{\alpha} q \alpha}\psi_\alpha\;,
\end{align}
\end{subequations}
which imply  $F_\alpha^{(B=q)}=0$. Next, there is a contribution originated from $B=\alpha, \bar{\alpha}$
\begin{eqnarray}
    F_\alpha ^{(B=\alpha,\bar{\alpha})} &=&\frac{\lambda}{\sqrt{2}}\left(\psi_{\bar{\alpha}}\wedge(\psi_\alpha\wedge\psi_{\bar{\alpha}}-vf_{\alpha \bar{\alpha} q}\psi_q)+i \psi_\alpha\wedge(\psi_{\bar{\alpha}}\wedge\psi_\alpha-vf_{\bar{\alpha}\alpha q}\psi_q)\right)\notag\\
    &=& \lambda\frac{\psi_{\bar{\alpha}}-i\psi_\alpha}{\sqrt{2}}\wedge\left(\psi_\alpha\wedge\psi_{\bar{\alpha}}-vf_{\alpha\bar{\alpha} q}\psi_q\right)\notag\\
    &=&\lambda\left[[\zeta_\alpha,\zeta_\alpha^\dagger]-v\alpha\cdot \psi,\zeta_\alpha\right]\;,
    \label{fromq}
\end{eqnarray}
where we used the property
\begin{equation}
    \psi_\alpha\wedge\psi_{\bar{\alpha}} = \left[\zeta_\alpha,\zeta_\alpha^\dagger\right]\;.
\end{equation}
Finally, we evaluate $ F_{\alpha}^{(B=\gamma,\bar{\gamma})} = P_\alpha + Q_\alpha$, where $P_\alpha$ ($Q_\alpha$) is the part without (with) explicit dependence on the structure constants. They are given by a sum over positive roots $\gamma \neq \alpha$
\begin{subequations}
\begin{align}
  P_\alpha &= \lambda\sum\limits_{\gamma \neq \alpha}\left( \psi_\gamma\wedge(\zeta_\alpha\wedge\psi_\gamma) + \psi_{\bar{\gamma}}\wedge(\zeta_\alpha\wedge\psi_{\bar{\gamma}})\right) \\
   Q_\alpha &= \frac{\lambda v}{\sqrt{2}}\sum\limits_{\gamma \neq \alpha}\left( f_{\alpha\gamma\bar{\delta}}\psi_\gamma \wedge \psi_{\bar{\delta}}-f_{\alpha\bar{\gamma}\delta}\psi_{\bar{\gamma}}\wedge \psi_\delta-if_{\bar{\alpha}\gamma\delta}\psi_\gamma\wedge\psi_\delta-if_{\bar{\alpha}\bar{\gamma}\bar{\delta}}\psi_{\bar{\gamma}}\wedge\psi_{\bar{\delta}}\right).
\end{align}
\end{subequations}
Using Eq. \eqref{CommutatorZetas}, we arrive at
\begin{eqnarray} 
    P_\alpha  &=&\lambda\sum\limits_{\gamma \neq \alpha} \left( \zeta_\gamma\wedge(\zeta_\alpha\wedge\zeta_\gamma^\dagger)+ \zeta_\gamma^\dagger\wedge(\zeta_\alpha\wedge\zeta_\gamma)\right) = \lambda\sum\limits_{\gamma \neq \alpha}\left(\left[ [\zeta_\gamma,\zeta_\gamma^\dagger],\zeta_\alpha\right]-2v \mathcal{N}_{\alpha,\gamma} [\zeta_\gamma^\dagger,\zeta_{\alpha+\gamma}]\right)\;.
\end{eqnarray}
On the other hand, by using Eqs. \eqref{commutation} and \eqref{Zetadefinition} it is possible to cast $Q_\alpha$ in the form 
\begin{eqnarray}
\label{LongBevaluation}
    Q_\alpha   &=&\lambda v \sum\limits_{\gamma \neq \alpha}\left( \mathcal{N}_{\alpha,\gamma}[\zeta_\gamma^\dagger,\zeta_{\alpha+\gamma}] +  \mathcal{N}_{\alpha,-\gamma} [\zeta_\gamma,\zeta_{\alpha-\gamma}]\right)
    \end{eqnarray} 
Let us analyze the term with  label $\alpha-\gamma$. Because $\gamma$ is a positive root, $\alpha-\gamma$ is not necessarily positive, so we cannot use Eq. \eqref{CommutatorZetas} right away. Instead, we shall split this term into two contributions: $\gamma=\gamma^+$ ($\gamma=\gamma^-$) such that $\alpha-\gamma^+$ ($\alpha-\gamma^-$) is a positive (negative) root. In the second case
\begin{equation}
    \lambda v \mathcal{N}_{\alpha,-\gamma^-}[\zeta_{\gamma^-},\zeta_{\alpha-\gamma^-}] = \lambda v \mathcal{N}_{\alpha,-\sigma-\alpha}[\zeta_{\sigma+\alpha},\zeta_{-\sigma}]=\lambda v \mathcal{N}_{\alpha,\sigma}\left[\zeta_\sigma^\dagger,\zeta_{\sigma+\alpha}\right] \;,
\end{equation}
where  $\sigma$ is a positive root that, when summed with $\alpha$, yields another positive root. This is precisely the condition on $\gamma$ in the first term of Eq. \eqref{LongBevaluation}. Therefore, 
\begin{equation}
    Q_\alpha = \lambda v\sum\limits_{\gamma \neq \alpha} 2 \mathcal{N}_{\alpha,\gamma}[\zeta_{\gamma}^\dagger,\zeta_{\alpha+\gamma}]+ \lambda v\sum\limits_{\gamma^+} \mathcal{N}_{\alpha,-\gamma^+}[\zeta_{\gamma^+},\zeta_{\alpha-\gamma^+}]\;,
\end{equation}
which together with the result for $P_\alpha$ yields
\begin{eqnarray}  
  F_{\alpha}^{(B=\gamma,\bar{\gamma})} 
  &=& \lambda\sum\limits_{\gamma \neq \alpha}\big[ [\zeta_\gamma,\zeta_\gamma^\dagger],\zeta_\alpha\big]+ \lambda v\sum\limits_{\gamma^+} \mathcal{N}_{\alpha,-\gamma^+} [\zeta_{\gamma^+},\zeta_{\alpha-\gamma^+}]  \;.
\end{eqnarray}
By the definition of $\gamma^+$, $\alpha-\gamma^+$ is positive so we can use Eq. \eqref{CommutatorZetas} once again to write
\begin{eqnarray}
     F_{\alpha}^{(B=\gamma,\bar{\gamma})} =&&\lambda\sum\limits_{\gamma \neq \alpha}\big[ [\zeta_\gamma,\zeta_\gamma^\dagger],\zeta_\alpha\big]+ \lambda v^2 \sum\limits_{\gamma^+}\mathcal{N}_{\alpha,-\gamma^+} \mathcal{N}_{\gamma^+,\alpha-\gamma^+}\zeta_{\alpha}\notag\\
     =&&\lambda\sum\limits_{\gamma \neq \alpha}\big[ [\zeta_\gamma,\zeta_\gamma^\dagger],\zeta_\alpha\big]-\lambda v^2\sum\limits_{\gamma^+}\mathcal{N}_{\alpha,-\gamma^+}^2 \zeta_{\alpha} \;.
     \label{ws}
\end{eqnarray}
To evaluate the sum over $\gamma^+$, we need to count how many roots are consistent with the $\alpha-\gamma^+>0$ condition. For this objective, we can use that $\alpha = \omega_I-\omega_J$ for some $I<J$.  Then, there are two cases 
\begin{eqnarray}
    \gamma^+ &=& \omega_I - \omega_l\;, I<l<J \Rightarrow J-I-1~~\text{ possibilities,}\notag\\
    \gamma^+ &=& \omega_l - \omega_J\;, I<l<J \Rightarrow J-I-1~~\text{ possibilities.}\notag
\end{eqnarray}
 Moreover, since $\mathcal{N}_{\alpha,-\gamma^+}^2 =  \frac{1}{2N}$ in both of these cases, we have  
\begin{equation}
    \sum\limits_{\gamma^+} \mathcal{N}_{\alpha,-\gamma^+}^2 = \frac{J-I-1}{N}\;.
\end{equation}
The sum of the $\mathcal{N}^2$-factors in Eq. \eqref{ws} can be rewritten as a sum of $(\alpha\cdot\gamma)$-factors: 
\begin{eqnarray}
    \sum\limits_{ \gamma \neq \alpha}\alpha\cdot\gamma = \frac{N+J-I-3}{2N} -\frac{N-J+I-1}{2N} = \sum\limits_{\gamma^+} \mathcal{N}_{\alpha,-\gamma^+}^2\;,
\end{eqnarray}
where we used a similar counting to determine  how many positive roots $\gamma$ different from $\alpha$ have $\alpha\cdot\gamma = \pm \frac{1}{2N}$. In addition, using the ansatz,
\begin{equation}
    \alpha\cdot\gamma \zeta_{\alpha} =  [\gamma\cdot T,\zeta_\alpha] \;,
\end{equation}
 so that   
\begin{equation}
    F_{\alpha}^{(B=\gamma,\bar{\gamma})} = \lambda\sum\limits_{\gamma \neq \alpha} \big[[\zeta_\gamma,\zeta_\gamma^\dagger]-v^2 \gamma \cdot T,\zeta_\alpha\big] \;.
\end{equation}
Finally, joining this result with the previous ones, namely $F_\alpha^{(B=q)}=0$ and  Eq. \eqref{fromq}, we get
\begin{equation}
    D^2\zeta_\alpha = \lambda\big[[\zeta_\alpha,\zeta_\alpha^\dagger]-v^2\alpha\cdot T,\zeta_\alpha\big]+\lambda\sum\limits_{\gamma \neq \alpha} \big[[\zeta_\gamma^\dagger,\zeta_\gamma]-v^2 \gamma \cdot T,\zeta_\alpha\big] = \lambda\sum\limits_{\alpha'>0}\big[v^2\alpha'\cdot T - [\zeta_{\alpha'},\zeta_{\alpha'}^\dagger],\zeta_\alpha\big]\;,
\end{equation}
which equals Eq. \eqref{BPSpsialphapredic} for $\lambda= g ^2$.
\section{Physical consequences}
 
In the previous sections, we showed that at $\mu^2 =0$, $\lambda= g^2$ the proposed vortex ansatz that closes the BPS equations provide static vortex solutions for the  $SU(N) \to Z(N)$ YMH model defined in Eq. \eqref{action}. From Eqs. \eqref{Zetadefinition}-\eqref{BPSeq}, the associated energy per unit-length is 
\begin{eqnarray} 
\label{BPSEnergy}
    \epsilon &=& \int d^2x\, \Big( \frac{1}{2}\langle B_3,B_3\rangle + \sum\limits_{\alpha>0}\langle D_i\zeta_\alpha^\dagger, D_i\zeta_\alpha\rangle+ V_{\rm H} (\psi) \Big)   \;,
\end{eqnarray}
where $d^2 x$ integrates over the transverse directions to the infinite string. 
Using Derrick's theorem in two dimensions, we can equate the potential energy of the Higgs field to
that of the gauge field, thus obtaining 
\begin{eqnarray}
\label{ModelScaling}
    \epsilon&=& \int d^2x\, \langle B_3,B_3\rangle - \langle \zeta_\alpha^\dagger,D^2\zeta_\alpha\rangle\notag\\
    &=&\int d^2x\, \langle B_3,B_3\rangle -\langle \zeta_\alpha^\dagger, D_-D_+\zeta_\alpha\rangle + g\langle\zeta_\alpha^\dagger,[B_3,\zeta_\alpha]\rangle = \int d^2x\, \langle B^3, B^3+g[\zeta_\alpha,\zeta_\alpha^\dagger]\rangle\notag\\
    &=&\int d^2x\, g v^2\langle B^3,2\delta\cdot T\rangle = gv^2\oint \, \langle A_i ,2\delta\cdot T\rangle  \, dx_i \;,
\end{eqnarray}
where $\delta$ is the sum of all positive roots and the last integral must be taken along a cicle with infinite radius. Recalling Eq. \eqref{AsymptoticGauge}, this implies that
\begin{gather}
\epsilon = 2\pi g v^2 \beta\cdot2\delta  \;.
\label{ener-law}
\end{gather}
 at the BPS point.
  In particular, note that the $k\text{-}{\rm A}$ string tension scales with the quadratic Casimir, as $\beta\cdot2\delta = \frac{N}{N+1} C_2({k\text{-}{\rm A}})$ in this case. This is the result we obtained in Ref.  \cite{oxmangustavo}. The new important physical consequence that we shall derive from Eq. \eqref{ener-law} is that for a general representation ${\rm D}(\cdot)$ with $N$-ality $k$, the asymptotic string tension satisfies
\begin{equation}
   \frac{\sigma({\rm D})}{\sigma({\rm F})} = \frac{ C_2({k\text{-}{\rm A}})}{C_2({\rm F})}\;,
   \label{casim-l}
\end{equation} 
which is one of the possible behaviors observed in lattice simulations. Indeed, when external monopole sources representing quarks are set apart, dynamical adjoint monopoles that represent valence gluons may be created around the sources, if this lowers the total energy.
As the adjoint representation has trivial $N$-ality,  the favored confining string at asymptotic distances will be the one with the lowest energy among those with the same $N$-ality ($k$) of ${\rm D}(\cdot)$. In what follows, we shall see that the smallest  $\beta\cdot2\delta $ factor is given by the $k\text{-}{\rm A}$ weight, which settles the behavior in Eq. \eqref{casim-l}.   

\begin{figure} 
    \centering
    \includegraphics[scale=0.6]{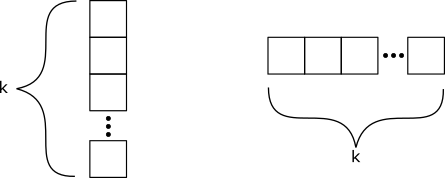}
    \caption{Young tableaux for the $k$-A (left) and $k$-S (right) representations.}
    \label{YTSymAntiSym}
\end{figure}

To prove this result, some Young Tableaux technology, useful to study the properties of the irreducible representations, is required. In this discussion, we shall closely follow the ideas in Ref. \cite{PhysRevD.64.105019}. 
A Young Tableau consists of a number of boxes organized according to the following rules:
\begin{enumerate}
    \item The maximum allowed number of boxes on a given column is $N-1$.
    \item The number of boxes in a given column ($n_i$) should be lower or equal than the number 
    in any column to the left. That is, $i > j \rightarrow n_i \leq n_j $.
    \item The number of boxes in a given row ($m_i$) should be lower or equal than the number in any row above. That is, $i > j \rightarrow m_i \leq m_j $.
\end{enumerate}
Every diagram drawn according to these rules corresponds to an irreducible representation of $SU(N)$. Many related properties can be easily identified in this language \cite{PhysRevD.64.105019}. The $N-$ality of a representation is simply given by the number of boxes of the Young Tableau, modulo $N$. The Dynkin indices $d_k$ of the highest weight $\Lambda$ satisfy \cite{PhysRevD.64.105019}\footnote{When $i=N-1$, we take $m_N=0$.} 
\begin{equation}
\Lambda= \sum\limits_{l=1}^{N-1}d_l \Lambda^{l\text{-A}}  \makebox[.5in]{,}    d_i=m_i-m_{i+1}\;.
\end{equation}
In general, when a box is moved from an upper to a lower row, an irrep. with more antisymmetries is obtained. For example, the Young tableau for the $k$-A ($k$-S) irrep. has one column (row) with $k$ boxes, as shown in Fig. \ref{YTSymAntiSym}.
For an irrep. with $N$-ality $k$, that is, a Young tableau with a total number of boxes of the form $k+nN$, the scaling factor can be written as
\begin{eqnarray}
\label{betadelta}
    \beta\cdot2\delta &=& \frac{N}{N+1}\sum\limits_{l=1}^{N-1}d_l\,l (N-l) =N(k+nN)-\frac{2N}{N+1}\sum\limits_{l=1}^{N-1} m_l\,l\;.
\end{eqnarray} 
Then, if a pair of irreps. ${\rm D}$ and ${\rm D}'$ with magnetic weights $\beta$ and $\beta'$, respectively, have the same $N$-ality $k$, we obtain
\begin{equation}
    \Delta\beta\cdot2\delta = \beta'\cdot2\delta-\beta\cdot2\delta = N^2\Delta n -\frac{2N}{N+1}\sum\limits_{l=1}^{N-1} \Delta m_l\, l\;,
\end{equation}
$\Delta m_l = m'_l-m_l$, $\Delta n = n'-n$, where the primed variables refer to ${\rm D}'$.
  Let us initially consider a pair of Young tableaux with the same number of boxes. 
\begin{figure}
    \centering
    \includegraphics[scale=0.6]{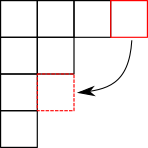}
    \caption{An example of transformation on a tableau that decreases the scaling factor $\beta\cdot2\delta$.}
    \label{YTMovingBoxes}
\end{figure}
If a  box is moved from an upper row $I$ to a lower row $J$ (see, for example, Fig. \ref{YTMovingBoxes}), we have $I<J$ and $\Delta m_J=-\Delta m_I =1$; consequently, $\Delta\beta\cdot2\delta =\frac{2N}{N+1}(I-J)<0$. This means that, for a given number of boxes $k+nN$, the tableau with smallest $\beta \cdot 2\delta$ is that in which the boxes are as lowered as possible.
\begin{figure}
    \centering
    \includegraphics[scale=0.6]{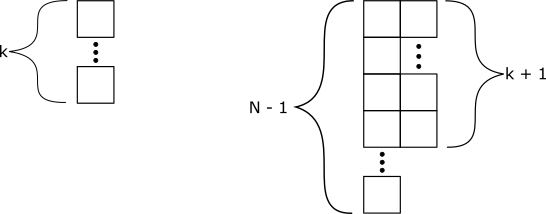}
    \caption{Fully antisymmetric Young tableau with $k$ (left) and $N+k$ (right) boxes.}
    \label{YT0to1}
\end{figure} 
Among these tableaux, we need to compare those having different $n$ but the same $N$-ality. As an initial example, let us begin by comparing the pair shown in Fig. \ref{YT0to1} and assume that the column of the first one is not completely full, i.e. $k\le N-2$. In this case, we see that 
\begin{equation}
    \Delta m_i=\begin{cases}
    2,\text{ if } i= k,\\
    1,\text{ otherwhise.}
    \end{cases}
\end{equation}

\begin{figure}
    \centering
    \includegraphics[scale=0.6]{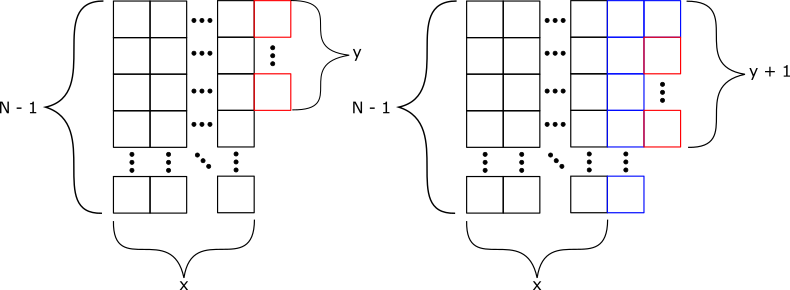} 
    \caption{Fully antisymmetric Young tableau with $k+nN$ (left) and $k+(n+1)N$ (right) boxes. There are $y$ boxes (in red) in the partly full column in the first tableau and $N$ boxes (in blue) were added in the second one (colors online).}
    \label{YTnton+1}
\end{figure}

Also, $\Delta n=1$ because we are comparing $k$ with $k+N$ boxes, in which case
\begin{eqnarray}
\label{DifferenceYT0to1}
    \Delta\beta\cdot2\delta &=& N^2\Delta n -\frac{2N}{N+1} \sum\limits_{l=1}^{N-1}\Delta m_l \,l  = \frac{2N}{N+1}(N-k)>0\;.
\end{eqnarray}
This means the scaling factor increases when we go from $k$ to $N+k$ boxes. This can  be readily extended to the general case depicted in Fig. \ref{YTnton+1}.
Because $\beta\cdot2\delta$ depends only on the difference of the number of boxes, the $x$ full columns in both diagrams can be disregarded for our purposes. The values of $x$ and $y$ are such that $y+x(N-1)=k+nN$.  In fact, the analysis of the relevant part of these two tableaux is completely analogous to that of Fig. \ref{YT0to1}, which leads to the same result of Eq. \eqref{DifferenceYT0to1} but with $y$ instead of $k$. Since $1\le y\le N-1$, the net difference continues to be positive. 

In summary, the smallest scaling factor within a given $N$-ality $k$ corresponds to the single column tableau on the left side of Fig. \ref{YT0to1}, namely, the one corresponding to the $k$-A representation. 
 
\section{Conclusions}

In this work, we were able to find a set of BPS equations which provide center string solutions for 
a Yang-Mills-Higgs model containing $N^2-1$ adjoint Higgs fields. This type of model can be thought of as an effective description for center-element averages over an ensemble of closed worldsurfaces and correlated worldlines.  It is strongly believed that these ensembles can capture the relevant quantum degrees of pure Yang-Mills theories in the infrared regime. In the ensemble, a center-element
 is generated every time a worldsurface links the Wilson loop. As this element depends on how the the quark representation realizes the center of $SU(N)$, this scenario is able to explain 
 the property of $N$-ality observed in the full Monte Carlo simulations of YM theory. 
 In the YMH model, $N$-ality is reflected in the $SU(N) \to Z(N)$ SSB pattern, while the information about the Wilson loop is manifested as a frustration in the effective gauge field sector. This in turn amounts to  represent the quark/antiquark in terms of monopole/antimonopole sources with charges in the given quark representation. As the distance between the quark and antiquark grows, to lower the total energy, the YMH model allows for the formation of dynamical adjoint monopoles localized around the sources (valence gluons).  These  objects cannot induce transitions that change the $N$-ality of the confining state, so that the asymptotic confining string will be the one with the lowest energy among those with the same $N$-ality. Here, we found the energy of an infinite string solution to the BPS equations in a general representation of $SU(N)$. We showed that the energy corresponding to the $k\text{-}{\rm A}$ representation is the lowest among all the quark representations with $N$-ality $k$. In other words, for widely separated quark/antiquark sources, 
the stable state is indeed given by the $k\text{-}{\rm A}$ string.\footnote{Of course, for the trivial $N$-ality $k=N$ (mod $N$) this corresponds to the string breaking.} This together with our previous result in Ref. \cite{oxmangustavo}, where the $k\text{-}{\rm A}$ string tension was shown to be proportional to the quadratic Casimir, completes the proof that the effective YMH model reproduces an asymptotic Casimir Law.

\appendix
\section{Cartan decomposition of $\mathfrak{su}(N)$} 
\label{appA}

Here, we  summarize the main properties of the $\mathfrak{su}(N)$ Lie algebra, as well as the conventions used throughout the paper. The construction of the Cartan-Weyl basis is initiated by defining a maximal commutative subspace,
whose generators $T_q$ satisfy 
\begin{equation}
    \left[T_q,T_p\right]=0\;,
\end{equation}
where $q,p=1,\dots,N-1$. The remaining basis elements are the so called root vectors $E_\alpha$, which diagonalize the adjoint action of $T_q$
\begin{equation}
    \left[T_q,E_\alpha\right] = \alpha\vert_q E_\alpha\;.
\end{equation}
The eigenvalues $\alpha\vert_q$ form an $(N-1)$-tuple  $\alpha=(\alpha\vert_1,\alpha\vert_2,\dots,\alpha\vert_{N-1})$ which is referred to as root. Since the dimensions of $\mathfrak{su}(N)$ and the Cartan subalgebra are, respectively, $N^2-1$ and $N-1$, there  are $N(N-1)$ root vectors. A well known result is that if $\alpha$ is a root, so is $-\alpha$. Moreover, the associated root vectors are related by
\begin{equation}
\label{Ealphadagger}
    E_{-\alpha} = E_{\alpha}^\dagger\;.
\end{equation}
We are considering the Cartan-Weyl basis $\{T_q,E_\alpha\}$ as orthonormal with respect to the product 
\begin{equation} 
    \langle A,B\rangle = {\rm Tr} \big( {\rm Ad}(A) {\rm Ad}(B) \big) \;,
\end{equation}
where ${\rm Ad}(\cdot)$ stands for the adjoint representation. In this case, we have
\begin{equation}
    \left[E_\alpha,E_{-\alpha}\right] =  \sum\limits_{q=1}^{N-1}\alpha_q T_q = \alpha\cdot T\;.
\end{equation}
In order to completely specify the commutation relations of root vectors, we need to address two roots that do not sum up to zero. These relations turn out to be
\begin{equation}
    \left[E_\alpha,E_{\alpha'}\right] = \mathcal{N}_{\alpha,\alpha'} E_{\alpha+\alpha'}\;,
\end{equation}
where  $\alpha'\ne -\alpha$ and $\mathcal{N}_{\alpha,\alpha'}$ vanishes when $\alpha+\alpha'$ is not a root.  With the normalization adopted, one can show that
\begin{equation}
    \mathcal{N}_{\alpha,\alpha'}^2 = \frac{1}{2N}
\end{equation}
whenever it does not vanish. These structure constant also have the property
\begin{equation}
    \mathcal{N}_{\alpha',\alpha}=\mathcal{N}_{-\alpha,-\alpha'}=-\mathcal{N}_{\alpha,\alpha'}\;.
\end{equation}
Moreover, if $\alpha,\alpha',\alpha''$ are roots that add up to zero, then
\begin{equation}
    \mathcal{N}_{\alpha,\alpha'}=\mathcal{N}_{\alpha'',\alpha}=\mathcal{N}_{\alpha',\alpha''}\;.
\end{equation}
The root vectors $E_\alpha$, which  live in the complexified Lie algebra, can be replaced by the hermitian generators $T_\alpha$ and $T_{\bar{\alpha}}$ in Eq. \eqref{hermit}. When using the latter as basis elements,  one must consider only positive roots $\alpha>0$ to avoid overcounting (for the notion of positiveness, see App. \ref{wei}). In this basis, the following commutation relations hold
\begin{gather}
\label{commutation}
[T_q,T_{\alpha} ] =  i\alpha|_qT_{\bar{\alpha}} \makebox[.5in]{,} 
[T_q,T_{\bar{\alpha}}] =-i\alpha|_qT_{\alpha} \makebox[.5in]{,} 
[T_{\alpha},T_{\bar{\alpha}} ] = i\alpha|_qT_q \;, \\
[T_{\alpha},T_{\beta} ] = \frac{i}{\sqrt{2}}\big( N_{\alpha,\beta}T_{\overline{\alpha+\beta}} + N_{\alpha,-\beta} T_{\overline{\alpha-\beta}} \big)\;, \\
[T_{\alpha},T_{\bar{\beta}} ] = -\frac{i}{\sqrt{2}}\big( N_{\alpha,\beta}T_{\alpha+\beta} - N_{\alpha,-\beta} T_{\alpha-\beta} \big)\;,\\  
[T_{\bar{\alpha}},T_{\bar{\beta}}] = -\frac{i}{\sqrt{2}} \big( N_{\alpha,\beta}T_{\overline{\alpha+\beta}} - N_{\alpha,-\beta} T_{\overline{\alpha-\beta}} \big)\;.
\end{gather}  
However, these relations remain true even for negative roots, recalling that the extended hermitian generators are not independent from their positive-root  counterparts, and satisfy
\begin{equation}
\label{NegativeRootsT}
    T_{-\alpha} = T_{\alpha}\;,\; T_{-\bar{\alpha}} = -T_{\bar{\alpha}}\;.
\end{equation}

\section{weights and representations of $\mathfrak{su}(N)$}
\label{wei}

A weight of an irreducible representation ${\rm D}$ of $\mathfrak{su}(N)$ is an $(N-1)$-tuple formed by the eigenvalues of a simultaneous eigenvector  of ${\rm D}(T_q),q=1,\dots,N-1$. Each irreducible representation, or irrep. for short, has its own set of weights. That corresponding to the fundamental representation has $N$ elements $ \omega_1,\omega_2,\dots,\omega_N $ constrained by
\begin{equation} 
    \omega_1+\omega_2+\dots +\omega_N = 0\;.
\end{equation}
The weights of the adjoint representation are the roots, as they are eigenvalues for the adjoint action $[T_q , \cdot]$. They can be expressed as the differences
\begin{equation} 
    \alpha = \omega_i-\omega_j\;,
\end{equation}
for some $i,j=1,\dots,N$, which is consistent with the previous counting of $N(N-1)$ roots. Some useful sums are
\begin{subequations}
\begin{align}
    \sum\limits_{i=1}^N \omega_i\vert_q \omega_i\vert_q = \frac{1}{2N}\delta_{qp} \makebox[.5in]{,}
    \sum\limits_{\alpha} \alpha\vert_q\alpha\vert_p = \delta_{qp}\;.
\end{align}
\end{subequations}

 A weight is said positive if its last nonvanishing component is positive. Consequently, a weight is greater than another if their difference is positive. In particular, given the set of weights of a given irrep., we can always determine the highest. For the fundamental representation, we choose the ordering convention
\begin{equation}
    \omega_1>\omega_2>\dots>\omega_N\;.
\end{equation}
Then, a root $\alpha=\omega_i-\omega_j$ is positive if and only if $i<j$. 

 Among the irreps. with $N$-ality $k$, we have the $k$-Symmetric ($k$-S) and $k$-Antisymmetric ($k$-A), $k=1,\dots,N-1$. They are constructed from the totally symmetric and antisymmetric decomposition of $k$ tensor products of the fundamental representation.  
The corresponding highest weights are given by\footnote{Notice that $\Lambda^{1\text{-}{\rm S}}=\Lambda^{1\text{-}{\rm A}}=\omega_1$.} 
\begin{equation}
    \Lambda^{k\text{-}{\rm S}} = k \omega_1\makebox[.5in]{,} \Lambda^{k \text{-}{\rm A}} = \sum\limits_{i=1}^k \omega_i\;.
\end{equation}
It is important to emphasize that the highest weight of any irrep. can always be written as a nonnegative integer linear combination of the $k$-Antisymmetric weights, which are called fundamental weights (not to be confused with the weights of the fundamental representation). The coefficients are called Dynkin numbers and there is a one-to-one  correspondence between irreps. and these combinations.

To end this quick review, the quadratic Casimir operator for a given representation ${\rm D}$ is 
\begin{equation}
    C_2({\rm D}) = \sum\limits_{A=1}^{N^2-1}{\rm D}(T_A){\rm D}(T_A)\;.
\end{equation} 
This operator commutes with every element of
 $\mathfrak{su}(N)$ and thus it is proportional to the identity matrix. The proportionality constant is known as the quadratic Casimir. For our choice of normalization, the quadratic Casimir for the fundamental, adjoint, $k$-S and $k$-A representations are, respectively,
\begin{equation}
    \frac{N^2-1}{2N^2}\makebox[.3in]{,} 1\makebox[.3in]{,} \frac{k(N+k)(N-1)}{2N^2}\makebox[.3in]{,}\frac{k(N-k)(N+1)}{2N^2}\;.
\end{equation}
Finally, for any irrep. ${\rm D}$, the quadratic Casimir can be expressed in the form
\begin{equation}
    C_2({\rm D}) = \lambda\cdot(\lambda +2\delta)\mathbb{I}_{\mathcal{D}}\;,
\end{equation}
where $\lambda$ is the highest weight and $\delta$ is the Weyl vector, given by half the sum of the positive roots.

%\bibliographystyle{unsrt}
%\bibliography{refs}

\end{document}